\documentclass[12pt]{iopart}

\usepackage[utf8]{inputenc}

\usepackage{amsmath}
\usepackage{amssymb}
\usepackage{latexsym}
\usepackage{mathrsfs}

\usepackage{boxedminipage}

\usepackage{graphicx}

\DeclareGraphicsExtensions{.eps, .jpg}

\begin{document}

\title{Core repulsion effects in alkali trimers}

\author{R. Guérout}
\address{Laboratoire Aim\'{e} Cotton, CNRS, B\^at. 505, Univ Paris-Sud,\\F-91405 Orsay Cedex, France}

\author{P. Sold\'an}
\address{Department of Chemical Physics and Optics, Faculty of Mathematics and Physics, Charles University in Prague, Ke Karlovu 3, CZ-12116 Prague 2, Czech Republic}

\author{M. Aymar}
\address{Laboratoire Aim\'{e} Cotton, CNRS, B\^at. 505, Univ Paris-Sud,\\F-91405 Orsay Cedex, France}

\author{J. Deiglmayr}
\address{Physikalisches Institut, Universit\"at Freiburg, Hermann-Herder-Strasse 3, 79104 Freiburg, Germany}

\author{O. Dulieu}
\address{Laboratoire Aim\'{e} Cotton, CNRS, B\^at. 505, Univ Paris-Sud,\\F-91405 Orsay Cedex, France}
\ead{olivier.dulieu@lac.u-psud.fr}

\date{\today}

\begin{abstract}
The present paper is related to a talk presented during the Symposium on Coherent Control and Ultracold Chemistry held during the Sixth Congress of the International Society for Theoretical Chemical Physics (ISTCP-VI, July 2008). The talk was entitled "Electronic structure properties of alkali dimers and trimers.
Prospects for alignment of ultracold molecules". Here we report on the electrostatic repulsion forces of the ionic cores at short separation, involved when the potential energy surfaces of alkali trimers are calculated with a quantum chemistry approach based on effective large-core potentials for ionic core  description. We demonstrate that such forces in the triatomic molecule can be obtained as the sum of three pairwise terms. We illustrate our results on the lowest electronic states of Cs$_3$, which are computed for the first time within a full configuration interaction based on a large Gaussian basis set. As a preliminary section, we also propose a brief introduction about the importance of alkali trimer systems in the context of cold and ultracold molecules.
\end{abstract}

\maketitle

\section{Introduction: a brief overview on cold molecules} 
\label{sec:introduction_a_brief_overview_on_cold_molecules}
	
The research field on translationally cold ($\approx 1$K and lower) and ultracold ($\approx 1$mK and lower) molecules is continuously expanding in many directions, involving an increasing number of groups throughout the world. The availability of gaseous samples of cold neutral molecules opens entirely new avenues for fascinating researches, offering the possibility to control all degrees of freedom of a quantum system. Molecular ions can also be trapped and cooled down inside atomic ion traps behaving like ionic crystals, where each ion is kept at a specific position with a residual motion equivalent to a temperature smaller than 1~K \cite{gerlich2008a}. Cold neutral molecules brought new perspectives in high-resolution molecular spectroscopy \cite{stwalley1999,jones2006,meerakker2005,gilijamse2007}. The expected accuracy of the envisioned measurements with ultracold molecules makes them appear as a promising class of quantum systems for precision measurements related to fundamental issues: the existence of the permanent electric dipole moment of the electron \cite{demille2000,hudson2002,kozlov2002,kawall2004} related to CP-parity violation \cite{hinds1980,cho1991,ravaine2005}, and the time-independence of the electron-to-nuclear and nuclear-to-nuclear mass ratios \cite{veldhoven2004,karr2005,schiller2005,chin2006,demille2008,zelevinsky2008}, or of the fine-structure constant \cite{hudson2006a}. Various proposals have been suggested for achieving quantum information devices \cite{demille2002,rabl2006,kotochigova2006,rabl2007} based on cold polar molecules (i.e. molecules exhibiting a permanent electric dipole moment). Elementary chemical reactions at very low temperatures could be manipulated by external electric or magnetic fields, therefore offering an extra flexibility for their control \cite{krems2005}. The large anisotropic interaction between cold polar molecules is also expected to give rise to quantum magnetism \cite{barnett2006}, and to novel quantum phases \cite{baranov2002,goral2002}. The achievement of quantum degeneracy with cold molecular gases \cite{jaksch2002,donley2002,herbig2003,jochim2003,greiner2003,zwierlein2003,regal2003,bourdel2004} together with the mastering of optical lattices \cite{bloch2005} built a fantastic bridge between condensed matter and dilute matter physics. Indeed, the smooth crossover between the Bose-Einstein condensation (BEC) of fermionic atomic pairs and the Bardeen-Cooper-Schrieffer (BCS) delocalized pairing of fermions related to superconductivity and superfluidity, has been observed experimentally \cite{bourdel2004,chin2004,bartenstein2004,zwierlein2004}.

Cold and ultracold neutral molecules can be formed along two paths:
	\begin{itemize}
	\item Existing molecules in their vibrational ground state can be slowed down and cooled by various means, such as interactions with external electric and magnetic fields, or by collision with surrounding particles (see for instance the review articles in refs.\cite{bethlem2003,krems2005,hutson2006}. This approach concerns a broad variety of small molecules (OH, NH, NO, SO$_2$, ND$_3$, CO, YbF, C$_7$H$_5$N, H$_2$CO, LiH, CaH,...), but is currently limited to the production of molecules with a translational temperature around 10~mK.
	\item Pairs of trapped ultracold atoms can be associated using laser fields (photoassociation, or PA) \cite{jones2006}, or time-varying magnetic fields (magneto- or Feshbach- association) \cite{kohler2006}. Translational temperatures as low as a few tens of microKelvins can be reached, and even lower when quantum degeneracy is achieved. In contrast to the previous case, these results are up to now limited to alkali diatomic molecules which are most often formed in highly-excited vibrational levels, i.e. with high internal energy.
	\end{itemize}

In the latter case, a breakthrough occurred in 2008, when the possibility to create ultracold bialkali molecules in the lowest vibrational level of their ground state or their lowest metastable triplet state, has been demonstrated. Caesium dimers have been created in their ground state $v=0$ level after a PA step followed by spontaneous emission and further vibrational pumping, using a sequence of shaped laser pulses \cite{viteau2008}. Dipolar molecules, namely LiCs, have been observed in their absolute ground state rovibrational level $v=0, J=0$ after PA and spontaneous emission \cite{deiglmayr2008a}. Samples close to the degeneracy regime of ultracold molecules in the $v=0$ level of their ground state have been observed for Cs$_2$ \cite{danzl2008,mark2009}, and KRb \cite{ni2008}, and in the $v=0$ level of their lowest triplet state for Rb$_2$ \cite{lang2008}, using the STIRAP technique (Stimulated Rapid Adiabatic Passage) to transfer the population from initial high-lying vibrational levels.
	

\section{Motivation of the present work} 
\label{sec:motivation_of_the_present_work}

The growing availability of ultracold samples of alkali diatomics brings the possibility to observe and study their interactions with neighboring ultracold atoms and molecules. In two independent - and almost simultaneous- experiments, inelastic rate constants for atom-molecule and molecule-molecule collisions have been extracted, using optically trapped Cs$_2$ molecules created by PA and spontaneous emission \cite{staanum2006,zahzam2006}. The rate constants have been found independent from the initially populated rovibrational level. A resonant feature observed in the loss rate after collisions between trapped ultracold Cs$_2$ molecules created by Feshbach association has been interpreted as Cs$_{2}$-Cs$_{2}$ bound states \cite{chin2005}. Inelastic atom-molecule collisions in a trapped sample of RbCs molecules have been studied for both RbCs-Cs and RbCs-Rb cases~\cite{hudson2008}. No systematic dependence on the internal state of the molecule has been probed within the experimental precision.
	
Theoretical knowledge of the electronic structure of alkali triatomic systems is strongly needed for theoretical dynamical studies, which would support these experiments. There has been a significant research activity concerning the theoretical study of the structure and dynamical properties of homonuclear alkali-metal trimers. Non-additive effects in spin-polarized alkali-metal trimers were studied by Sold\'{a}n \textit{et.al.} \cite{soldan2003} and the three-dimensional potential energy surfaces for the lowest quartet states were constructed for Li$_{3}$~\cite{colavecchia2003a,cvitas2007}, Na$_{3}$~\cite{higgins2000,simoni2009}, K$_{3}$ \cite{quemener2005} and Rb$_{3}$ \cite{soldan2008u}. These were then used to study the corresponding spin-polarized reactive atom-dimer collisions at very low temperatures \cite{cvitas2007,quemener2005,soldan2002,quemener2004,cvitas2005a,cvitas2005,quemener2007,li2008a,li2008b,li2008c}. General trends for the lowest quartet state of homonuclear alkali trimers - correlated to three spin-polarized alkali atoms - have been described by Hutson and Sold\'an \cite{hutson2007}. One of the main characteristic of these systems is the importance of non-additive three-body forces, which are predicted to be large especially close to the equilibrium geometries \cite{soldan2003}. Much less has been done in the case of heteronuclear alkali-metal trimers. One of us has actually started with a systematic study the lowest quartet states of Li$_{2}$A mixed systems (with A=Na, K, Rb, Cs) \cite{soldan2008}. He concluded that the single-reference coupled-cluster approach, which has been successfully used for homonuclear alkali-metal trimers \cite{quemener2005,soldan2008}, would be for various reasons very difficult to employ for calculations of the potential energy surfaces of heteronuclear alkali-metal trimers, and that an alternative approach to the problem should be sought.
	
In the present paper, we present such an alternative approach. We extend our previous works on effective two-electron diatomic molecules like alkali dimers \cite{aymar2005,aymar2006,deiglmayr2008} and alkali hydrides \cite{aymar2009} to alkali trimers. We propose a preliminary study of the potential energy surfaces of the heaviest alkali trimer, Cs$_3$, modeled as an effective three-electron molecule. Our approach is based on large-core effective core potentials (ECP) with core polarization potentials (CPP). We focused our work on two aspects: (i) the estimation of repulsion effects between the three Cs$^+$ ionic cores at short distances within the present ECP+CPP approach; (ii) the comparison of the results with the results obtained by the MOLPRO package for {\it ab-initio} calculations.
	

\section{Method of calculation} 
\label{sec:method_of_calculation}
	
	\subsection{\emph{Ab initio} methods and basis sets} 
	\label{sub:ab_initio_methods_and_basis_sets}
		
The potential energies were calculated making use of the CIPSI package (Configuration Interaction by Perturbation of a multiconfigurational wave function Selected Iteratively)~\cite{huron1973}. As in our previous studies on alkali dimers, the alkali atom A is described by an $\ell$-dependent Effective Core Potential (ECP) for the ionic core A$^+$ \cite{durand1974,durand1975} including effective core polarization potential (CPP) terms \cite{muller1984,foucrault1992}, and by a large set of uncontracted Gaussian functions for the valence electron. The atom is modeled as an effective one-electron system, permitting us to perform Full Configuration Interaction (FCI) calculations for any alkali dimer or trimer, considered as an effective two-electron or three-electron system, respectively.

As the distance between different cores becomes smaller than the equilibrium distance of the system, it is well known that their electrostatic repulsion is not accounted for by this effective large-core potential. At the same time, the addition of the CPP term leads to an non-physical attractive behavior at short distances. For diatomic molecules, an empirical repulsion term is usually added to the calculation (see e.g. Refs.~\cite{spiegelmann1989,pavolini1989,magnier1993,magnier1996}). If this short-range repulsive interaction in the triatomic molecule can be treated as the sum of pairwise interaction terms, computation of the potential surfaces would be greatly simplified. In order to check if such an approximation could be justified, we calculated the core-core repulsion at the Hartree-Fock level making use of the MOLPRO 2006.1 Quantum Chemistry Package~\cite{MOLPRO_brief}. For lithium and sodium, the all-electron correlation-consistent polarized core-valence quadruple-$\zeta$ cc-pCVQZ basis sets~\cite{iron2003} were used. For potassium, rubidium, and caesium, the small-core scalar relativistic effective core potentials ECP10MDF, ECP28MDF, and ECP46MDF, respectively, together with the corresponding uncontracted valence basis sets~\cite{lim2005a} were employed.

For the caesium dimer and trimer, the counterpoise-corrected dimer interaction energies were optimized using an algorithm implemented in MOLPRO. In this algorithm the interaction energies were calculated at the coupled-clusters level making use of a single-reference restricted open-shell variant \cite{knowles1993} of the coupled cluster method \cite{cizek1966} with single, double and non-iterative triple excitations [RCCSD(T)]. The small-core scalar relativistic effective core potentials ECP46MDF was employed together with its uncontracted valence basis set, which was augmented by one set of diffuse function in an even-tempered manner (aug-ECP46MDF). All electrons from the ``outer-core'' $5s5p$ orbitals were included in the RCSSD(T) calculations. The full counterpoise correction of Boys and Bernardi \cite{boys1970} was applied to all the interaction energies in order to compensate for the basis set superposition errors.
		

	\subsection{Core Polarization Potential for a triatomic molecule} 
	\label{sub:core_polarization_potential}
	
The core polarization potential (CPP) is given by~\cite{muller1984}:

		\begin{equation}
		  \label{eq:VCPP}
		  V_{CPP}=-\frac{1}{2} \sum_{c} \alpha_{c}\,\mathbf{f}_{c}\cdot\mathbf{f}_{c}
		\end{equation}

The electric field $\mathbf{f}_{c}$ is the one produced at point $\mathbf{r}_{c}$ by all the other electrons and the ionic cores with static dipole polarizability $\alpha_{c}$:

		\begin{equation}
		  \label{eq:fc}
		  \mathbf{f}_{c}=\sum_{i}\frac{\mathbf{r}_{ci}}{r_{ci}^{3}}h(r_{ci},\rho_{c})\
			-\sum_{c^{\prime}\neq c}\frac{\mathbf{R}_{cc^{\prime}}}{R_{cc^{\prime}}^{3}}Z_{c^{\prime}}
		\end{equation}

where $\mathbf{r}_{ci}$ is the position vector from core $c$ to electron $i$, $\mathbf{R}_{cc^{\prime}}$ the one between cores $c$ and $c^{\prime}$. The cutoff function $h(r,\rho)$  ensures the convergence of the integral over electronic coordinates. The $V_{CPP}$ term contains a purely geometrical	part $v_{nn}$ depending only on the relative positions of the nuclei:

		\begin{equation}
		  \label{eq:vnn}
		  v_{nn}=-\frac{1}{2}\sum_{c}\alpha_{c}\sum_{c^{\prime},c^{\prime\prime}\neq c}\
			\frac{\mathbf{R}_{cc^{\prime}}\cdot \mathbf{R}_{cc^{\prime\prime}}}{R_{cc^{\prime}}^{3}\
			R_{cc^{\prime\prime}}^{3}}Z_{c^{\prime}}Z_{c^{\prime\prime}}
		\end{equation}

For a diatomic molecule, $v_{nn}$ is given by the familiar formula: $v_{nn}=-\frac{Z_{2}^{2}\alpha_{1}+Z_{1}^{2}\alpha_{2}}{2 R^{4}}$. For a triatomic system, we have:
		
		\begin{equation}
			\label{eq:CPP_tri}
			\begin{split}
				&v_{nn}=-\frac{Z_{2}^{2}\alpha_{1}+Z_{1}^{2}\alpha_{2}}{2 R_{12}^{4}}-\
				\frac{Z_{3}^{2}\alpha_{1}+Z_{1}^{2}\alpha_{3}}{2 R_{13}^{4}}\
				-\frac{Z_{3}^{2}\alpha_{2}+Z_{2}^{2}\alpha_{3}}{2 R_{23}^{4}}\\
				&-\frac{Z_{2}Z_{3}\alpha_{1}\,\mathbf{R}_{12}\cdot\mathbf{R}_{13}}{R_{12}^{3}R_{13}^{3}}\
				-\frac{Z_{1}Z_{3}\alpha_{2}\,\mathbf{R}_{12}\cdot\mathbf{R}_{23}}{R_{12}^{3}R_{23}^{3}}\
				-\frac{Z_{1}Z_{2}\alpha_{3}\,\mathbf{R}_{13}\cdot\mathbf{R}_{23}}{R_{13}^{3}R_{23}^{3}}
			\end{split}
		\end{equation}

In those formulas, $Z_{i}$ and $\alpha_{i}$ are, respectively, the net charge and static polarizability of ion core $i$. This term is independent of the \emph{ab initio} calculation itself and is added \emph{a posteriori}.
	
	

\section{Volume effect for triatomic systems} 
\label{sec:volume_effect_for_triatomic_systems}

The part of the interaction between two nuclei other than the point charge repulsion has been dubbed “volume effect” by Jeung in his 1997 paper on alkali diatomics~\cite{jeung1997}. Here, we are interested in this volume effect for alkali triatomic systems. More precisely, we investigate whether it can be reasonably approximated as a sum of two-body terms.
	
\begin{figure}[htbp]
 			\begin{center}
 				\includegraphics[width=15cm]{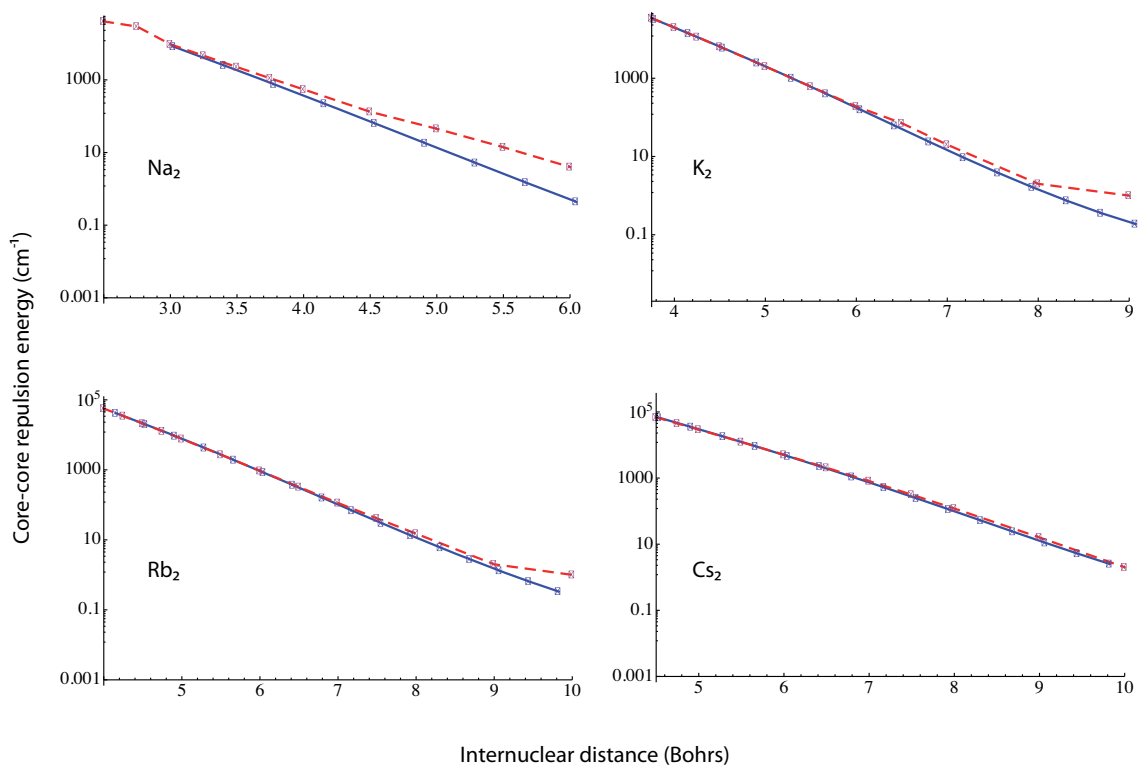}
 			\end{center}
 			\caption{Two-body core repulsion energies for alkali dimers. Blue : our work. Red dashed : Ref.~\cite{jeung1997}. Note the logarithmic scale used for the potential energy axis.}
 			\label{fig:compJeung}
 		\end{figure}

The two-body repulsion energies $V_{cc}$ are calculated for all the alkali homonuclear diatomic molecules. The three-body repulsion energies $V_{ccc}$ are calculated for particular geometries, namely for D$_{\infty h}$ and D$_{3h}$ symmetries, to provide a representative investigation. Calculations ar done at the Hartree-Fock level of theory starting with the orbitals of the free atoms. Those orbitals are kept frozen as the internuclear distance is shortened. In this respect, our calculations for the two-body term are similar to those of the “A” column from the tables of ref.~\cite{jeung1997}. This comparison is shown in Figure~\ref{fig:compJeung} in logarithmic scale. Apart from a slight discrepancy in the case of Na$_{2}$, there is a good agreement between both calculations, and it is seen that the two-body core repulsion term could be well fitted by an exponential form, as already stated in ref.\cite{pavolini1989}.

\begin{figure}[htbp]
		\begin{center}
			\includegraphics[width=10cm]{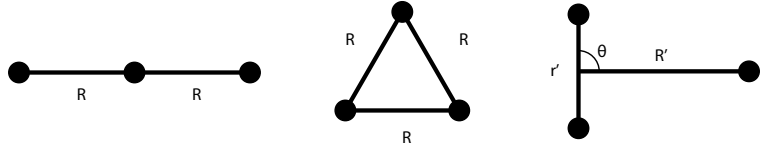}
		\end{center}
		\caption{Linear (D$_{\infty h}$) and equilateral (D$_{3h}$) geometries at which the three-body core repulsion term have been calculated. The Jacobi coordinates $\{r',R',\theta\}$ used in the calculation of the potential energy surfaces of Cs$_{3}$ are also presented.}
		\label{fig:geom}
	\end{figure}

\begin{figure}[htbp]
		\begin{center}
			\includegraphics[width=15cm]{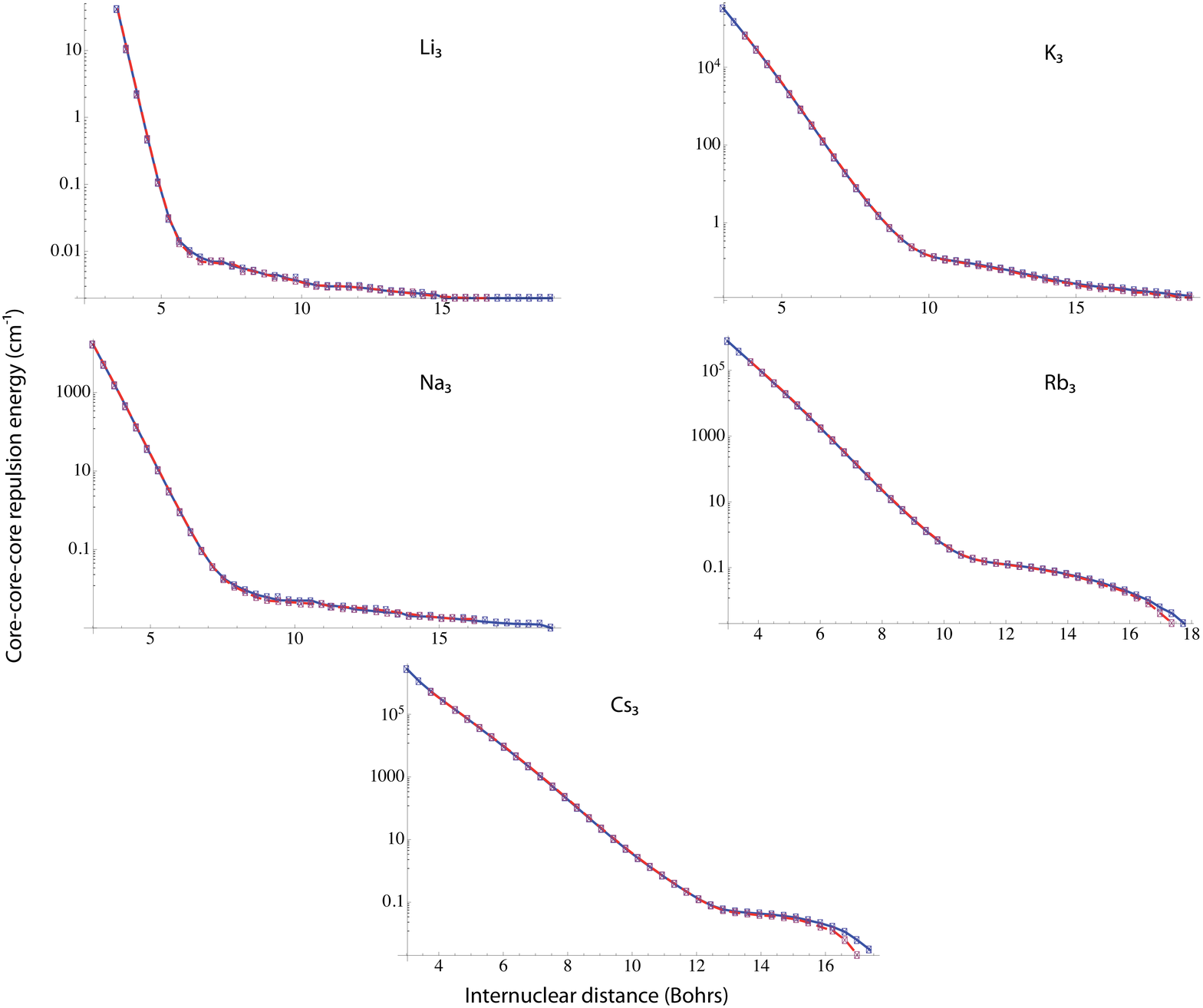}
		\end{center}
		\caption{Additivity of the three-body core repulsion term for linear geometries.
		Blue : Three-body term $V_{ccc}^{lin}(R)$. Red dashed :  sum of the relevant two-body terms $2 V_{cc}(R)+V_{cc}(2R)$ (see text).}
		\label{fig:addLin}
	\end{figure}

\begin{figure}[htbp]
		\begin{center}
			\includegraphics[width=15cm]{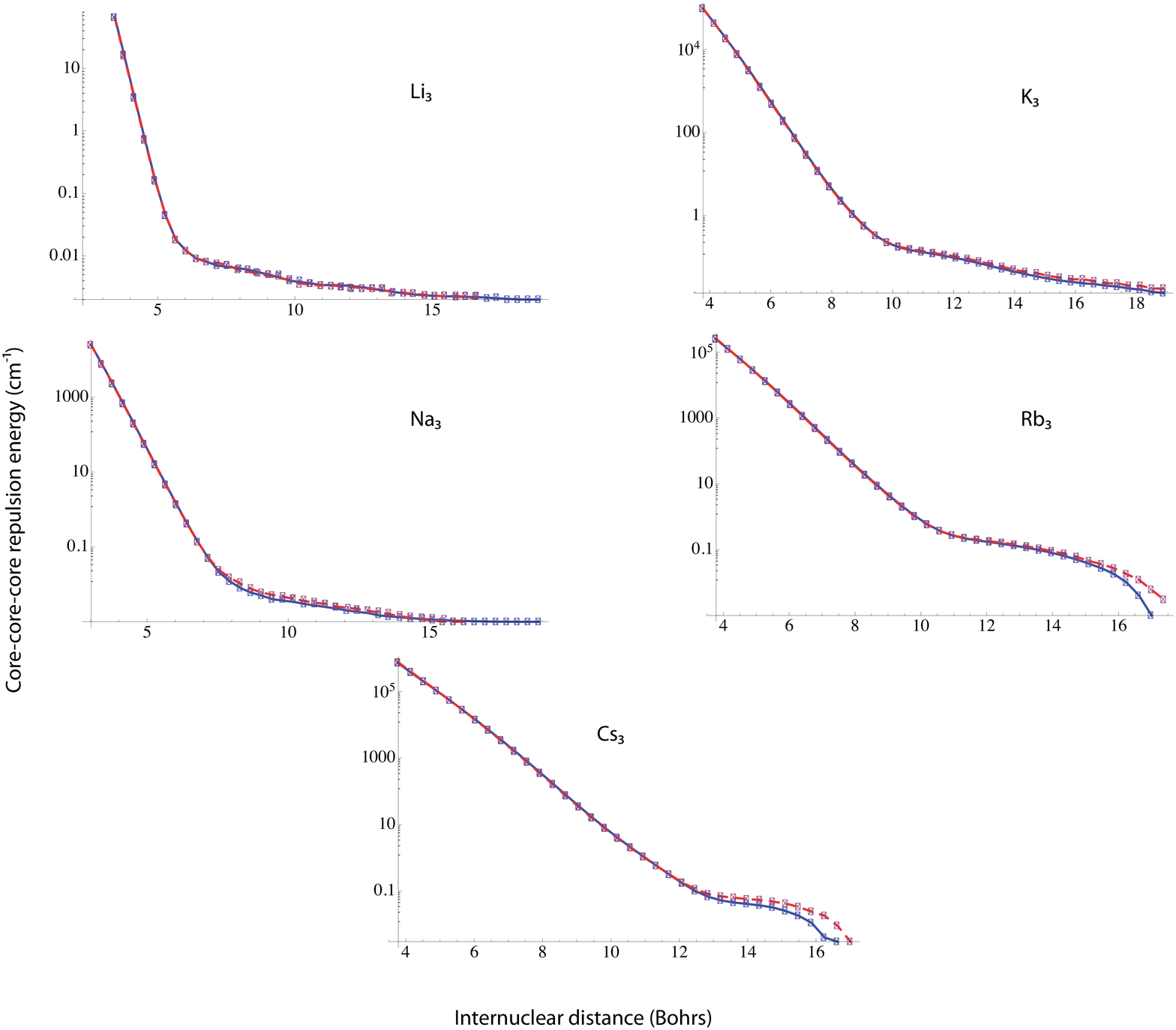}
		\end{center}
		\caption{Additivity of the three-body core repulsion term for equilateral geometries.
		Blue : Three-body term $V_{ccc}^{eq}(R)$. Red dashed : sum of the relevant two-body terms $3 V^{(2)}(R)$ (see text).}
		\label{fig:addEqui}
	\end{figure}

Three-body core repulsion energies have been calculated for all homonuclear alkali trimers at linear symmetric and equilateral geometries according to the schemes of Figure~\ref{fig:geom}. In Figure~\ref{fig:addLin}, we compare the three-body core repulsion term for linear geometries $V_{ccc}^{lin}(R)$ (the internuclear distance $R$ being defined in Figure~\ref{fig:geom}), with the sum of two-body term $2 V_{cc}(R)+V_{cc}(2R)$. In the same manner, we compare in Figure~\ref{fig:addEqui} the three-body repulsion term for equilateral geometries $V_{ccc}^{eq}(R)$ to $3 V^{(2)}(R)$. In both cases, it is seen that the three-body core repulsion term is well described by a pairwise additive lemma, up to the region where it becomes negligible. Therefore, just like for alkali dimers, the $V_{ccc}$ term can be added {\it a posteriori} to the potential surface calculations, whatever the chosen grid is for them.
	
\newpage
\section{Preliminary application to caesium trimer} 
\label{sec:preliminary_application_to_caesium_trimer}
	
The inclusion of the three-body core repulsion term is considered in preliminary calculations concerning the $^{4}A_{2}^{\prime}$ state of Cs$_{3}$ which correlates to three spin-polarized caesium atoms. The used basis set is kept relatively small to maintain reasonable size for the FCI, reaching $152\,532$ Slater determinants in C$_{s}$ symmetry in which the calculation is performed. The basis set $(5s3p4d/4s3p3d)$ is detailed in Table~\ref{tab:basisSet}. We use $\ell$-dependent CPP's~\cite{foucrault1992} which allow us to adjust the calculated atomic energies to the experimental ones for the $6s\,^{2}S$, $6p\,^{2}P$ and $5d\,\,^{2}D$ atomic states of caesium. This is achieved by tuning the cutoff radii $\rho_{\ell}$ of Eq.~\ref{eq:fc} to the values displayed in Table~\ref{tab:cutoffRadii}.

	\begin{table}
		\begin{center}
			\begin{tabular}{ccc}
				\hline
				Angular momentum &  Exponent &  Contraction coefficients\\
				\hline
				s &  0.347926  &   0.411589\\
				s &  0.239900  &  -0.682422\\
				s &  0.050502  &   1.\\
				s &  0.036900  &   1.\\
				s &  0.00515   &   1.\\
				p &  0.1837    &   1.\\
				p &  0.0655    &   1.\\
				p &  0.0162    &   1.\\
				d &  0.2106894 &   0.18965\\
				d &  0.065471  &   0.22724\\
				d &  0.021948  &   1.\\
				d &  0.011200  &   1.\\
				\hline
			\end{tabular}
		\end{center}
		\caption{Gaussian basis set used on each caesium atom in the ECP+CPP-FCI approach.}
		\label{tab:basisSet}
	\end{table}
	
	\begin{table}
		\begin{center}
			\begin{tabular}{cc}
				\hline
				$\ell$ & $\rho_{\ell}$\\
				\hline
				s & 2.6248\\
				p & 1.87\\
				d & 2.8111\\
				\hline
			\end{tabular}
		\end{center}
		\caption{$\ell$-dependent cutoff radii $\rho_{\ell}$ (in units of $a_{0}$) used in the core polarization potential. The polarization for Cs$^{+}$ ionic core of 16.33 $a_{0}^{3}$ is taken from Ref~\cite{coker1976}.}
		\label{tab:cutoffRadii}
	\end{table}
	
	\subsection{The triplet state of Cs$_{2}$} 
	\label{sub:the_triplet_state_of_cs2}
	
\begin{figure}[htbp]
		\begin{center}
			\includegraphics[width=13cm]{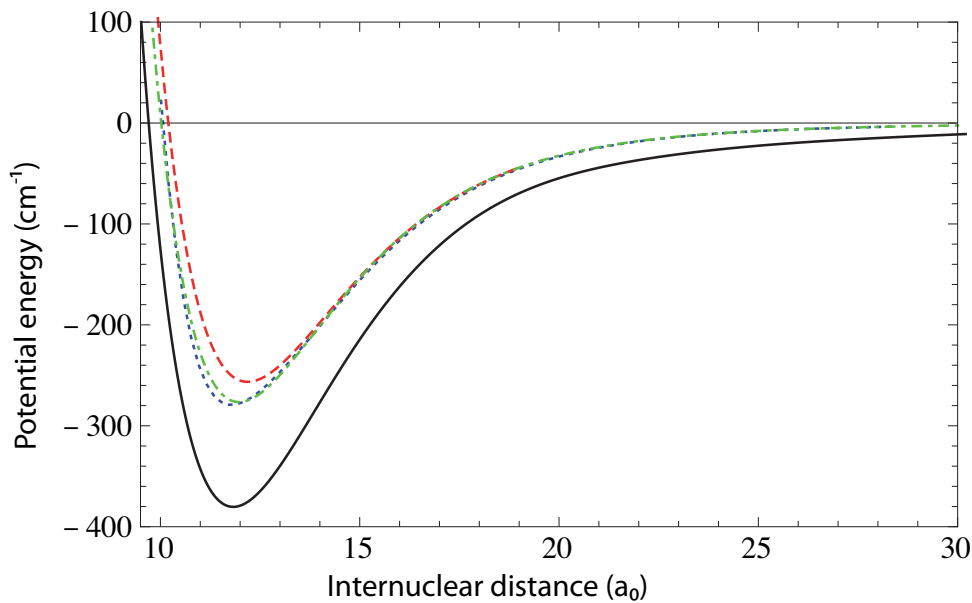}
		\end{center}
		\caption{Calculated potential energy curves for the lowest $^{3}\Sigma_{u}^{+}$ state of Cs$_{2}$. Black plain curve : ECP+CPP-FCI calculation with the basis detailed in Table~\ref{tab:basisSet}. Red dashed curve : obtained at the RCCSD(T)/aug-ECP46MDF level. Green chain dotted curve : ECP+CPP full CI result obtained with basis set “A” from Ref.~\cite{aymar2005} (see the discussion in Section~\ref{sec:conclusion}). Blue dotted curve : Multiparameter Morse Long Range fitting to spectroscopic data measured by Xie \emph{et al}~\cite{xie2009}.}
		\label{fig:cs2Triplet}
	\end{figure}

With these basis set and cut-off radii, we first calculate the lowest $^{3}\Sigma_{u}^{+}$ potential energy curve for Cs$_{2}$ dissociating into Cs($6s$)+Cs($6s$) in the framework of our ECP+CC-FCI method. We found an equilibrium internuclear distance R$_{e}=11.82\,a_{0}$ and a well depth of D$_{e}=380.6\,\text{cm}^{-1}$. These numbers can be compared to our optimized results obtained at the RCCSD(T)/aug-ECP46MDF level R$_{e}=12.19\,a_{0}$ and D$_{e}=256.4\,\text{cm}^{-1}$ and the results from Ref.~\cite{soldan2003,hutson2007}, who obtained R$_{e}=12.44\,a_{0}$ and D$_{e}=246.8\,\text{cm}^{-1}$ at the RCCSD(T)/ECP46MWB level of theory. The MOLPRO calculation yields a potential well which is deeper by is a more than $140\,\text{cm}^{-1}$ than the one of the ECP+CC-FCI calculation. Apparently, the small-core ECP in combination with a rich valence basis set in the RCCSD(T) calculations provides more realistic description of the interaction energy than the large-core ECP in combination with a smaller valence basis set used in the FCI calculations (see also the discussion in Section \ref{sec:conclusion}). The potential energy well of the triplet state results from the competition between the exchange energy at short distances and the attractive dispersion forces at long distances and is very sensitive to the quality of the basis. For this $^{3}\Sigma_{u}^{+}$ curve, the inclusion of the two-body core repulsion term does not change the characteristics of the well but ensures a correct exponential behavior at short internuclear distances (Figure~\ref{fig:cs2Triplet}).

	
	\subsection{Quartet state of Cs$_{3}$} 
	\label{sub:quartet_state}
	
We then calculate the potential energy surface of the lowest quartet state of Cs$_{3}$. Calculations are carried out in Jacobi coordinates $\{r',R',\theta\}$ (depicted in Figure~\ref{fig:geom}) with the angle $\theta$ fixed at $\pi/2$ in order to explore C$_{2v}$ nuclear geometries, as a representative geometry. We have computed 984 points on a grid where $4.4\,a_{0}<r'<40\,a_{0}$ and $4\,a_{0}<R'<40\,a_{0}$. The minimum of the surface is found at a D$_{3h}$ geometry with a bond length of $R'_{e}=10.72\,a_{0}$ and a well depth of $D_{e}=1518\,\text{cm}^{-1}$ with the inclusion of the pairwise three-body core repulsion term, which ensures a realistic short-range repulsive wall. These numbers can be compared to our optimization results obtained at the RCCSD(T)/aug-ECP46MDF level $R'_{e}=11.18\,a_{0}$ and D$_{e}=1208\,\text{cm}^{-1}$ and to the results from Ref.~\cite{soldan2003,hutson2007}, who obtained $R'_{e}=11.33\,a_{0}$ and D$_{e}=1139\,\text{cm}^{-1}$ at the RCCSD(T)/ECP46MWB level of theory. It is worth noting that as the minimum of the quartet potential well is located at shorter distances than the minimum of the Cs$_2$ triplet state, the $V_{ccc}$ term indeed contributes to the depth of the well, and not only in the region of the repulsive wall. However, the magnitude of $V_{ccc}$ is still small, as the equilibrium internuclear distance is $R'_{e}=10.7\,a_{0}$ and the well depth is $D_{e}=1522\,\text{cm}^{-1}$ if we neglect it. As expected from the diatomic calculations above, we obtain a difference of about $300\,\text{cm}^{-1}$ on the well depth of the quartet state of Cs$_{3}$ compared to the MOLPRO calculations. This confirms the limited quality of our basis set at the current level of our computations.


	\subsection{Doublet states of Cs$_{3}$} 
	\label{sub:doublet_states}
	
\begin{figure}[htbp]
		\begin{center}
			\includegraphics[width=13cm]{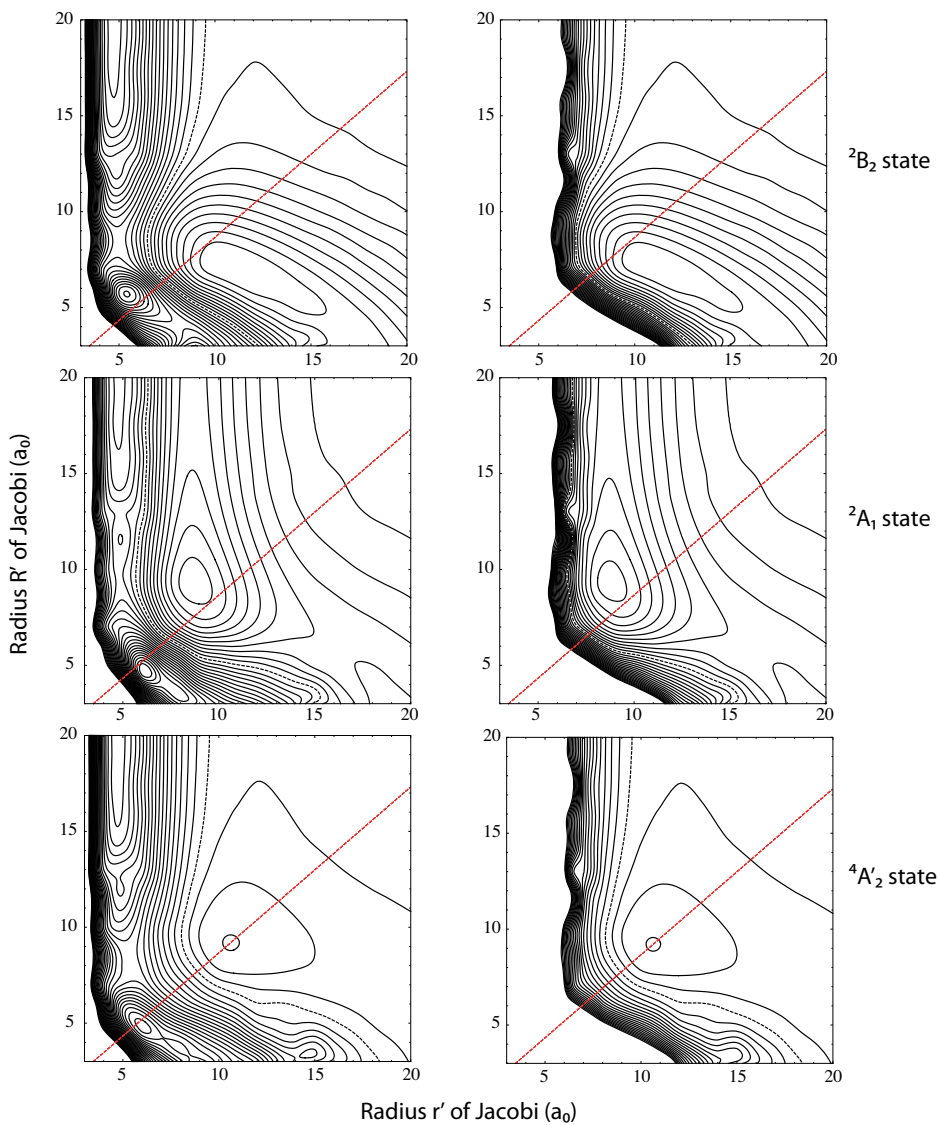}
		\end{center}
		\caption{Lowest three calculated surfaces for Cs$_{3}$ in Jacobi coordinates with fixed $\theta=\pi/2$. Left (resp. right) column : without (resp. with) the inclusion of the cores repulsion term $V_{ccc}$. The red dashed curve locates the equilateral D$_{3h}$ geometry $R'=\frac{\sqrt{3}}{2}r'$. Contours are distant of $500\,\text{cm}^{-1}$ and the dashed contour is the energy of the dissociation $\text{Cs}(^{2}S)+\text{Cs}(^{2}S)+\text{Cs}(^{2}S)$.}
		\label{fig:cs3Surfaces}
	\end{figure}

The three lowest potential surfaces for Cs$_{3}$ are presented in Figure~\ref{fig:cs3Surfaces}. The $^{2}B_{2}$ and $^{2}A_{1}$ states are the two components of a $^{2}E^{\prime}$ state at D$_{3h}$ symmetry subjected to Jahn-Teller effect. The left (resp. right) column shows the surfaces without (resp. with) the addition of the core repulsion term. Note the strongly unphysical attractive behavior in the region $r'<5\,a_{0}$, which is removed when $V_{ccc}$ is added. The global ground state $^{2}B_{2}$ is characterized by a well depth of $5437.1\,\text{cm}^{-1}$ at the geometry $\{r'=10.58\, a_{0},R'=7.36\, a_{0}\}$ with the inclusion of the core repulsion term. This changes to a well depth of $5489\,\text{cm}^{-1}$ at  $\{r'=10.57\, a_{0},R'=7.3\, a_{0}\}$ without the core repulsion term. Once again, the effect of $V_{ccc}$ is larger than in the dimer, as the equlibrium distance is shorter. The other component $^{2}A_{1}$ is characterized by a well depth of $5258.2\,\text{cm}^{-1}$ at $\{r'=8.75\, a_{0},R'=9.28\, a_{0}\}$ with the core repulsion term ($5312\,\text{cm}^{-1}$ at $\{r'=8.66\, a_{0},R'=9.26\, a_{0}\}$ without it).

	
\begin{figure}[htbp]
		\begin{center}
			\includegraphics[width=13cm]{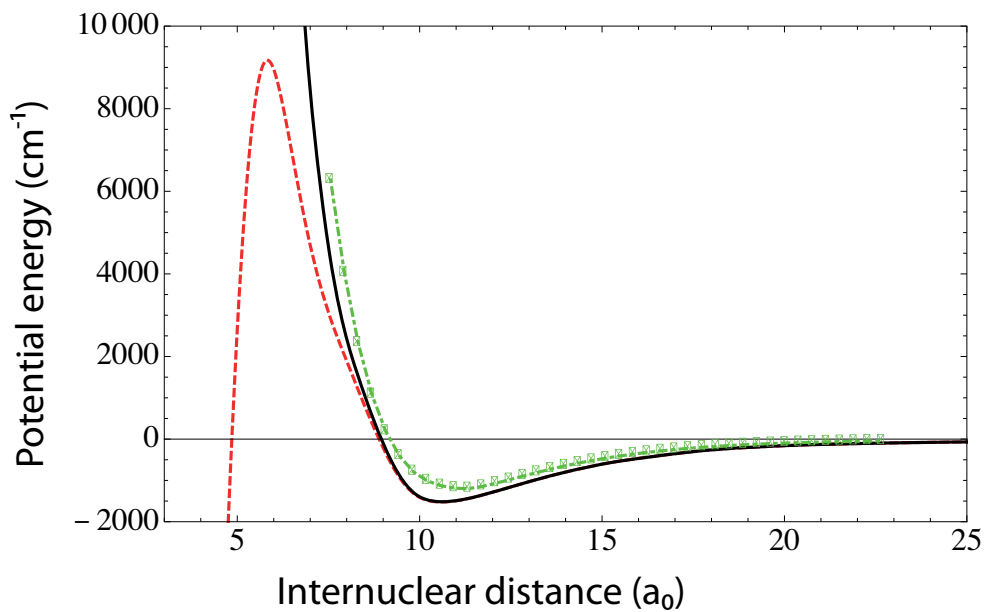}
		\end{center}
		\caption{Cut through the D$_{3h}$ geometry for the lowest quartet states of Cs$_{3}$. Plain black (resp. red dashed) curve : with (resp. without) the core repulsion term $V_{ccc}^{eq}(R)$. Green chain dotted curve : quartet state on the RCCSD(T)/aug-ECP46MDF level of theory (see the discussion in section \ref{sec:conclusion}.}
		\label{fig:cs3CurvesQuartet}
	\end{figure}

\begin{figure}[htbp]
		\begin{center}
			\includegraphics[width=13cm]{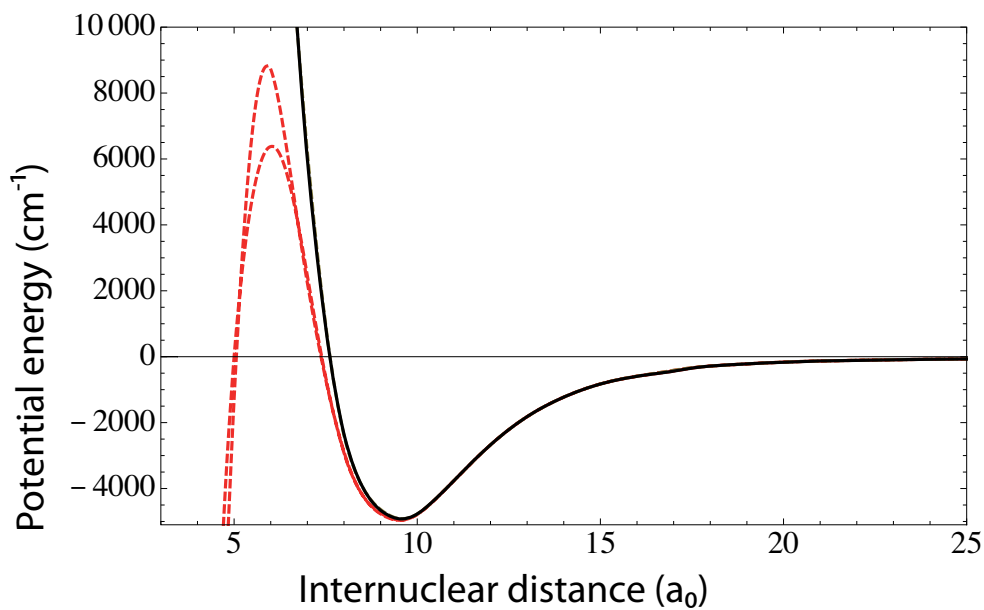}
		\end{center}
		\caption{Cut through the D$_{3h}$ geometry for the lowest doublet states of Cs$_{3}$. Plain black (resp. red dashed) curve : with (resp. without) the core repulsion term $V_{ccc}^{eq}(R)$. Both components of the $^{2}E'$ state are represented.}
		\label{fig:cs3CurvesDoublet}
	\end{figure}

We present in Figures~~\ref{fig:cs3CurvesQuartet} and \ref{fig:cs3CurvesDoublet} the cut through the D$_{3h}$ geometry of the previous surfaces. We recall that in the D$_{3h}$ symmetry point group, the two doublet states $^{2}B_{2}$ and $^{2}A_{1}$ discussed before correlates to the two components of a doubly degenerate $^{2}E'$ state. We clearly see in these figures the abrupt attractive behavior which occurs at short distances and that the addition of the core repulsion term gives us back a realistic repulsive wall of potential.
	
\section{Future prospects} 
\label{sec:conclusion}

Preliminary potential energy surfaces for doublet and quartet states of the Cs$_{3}$ molecule have been computed in the framework of a ECP+CPP-FCI quantum chemistry approach. The core repulsion interaction has been taken into account as an additional empirical correction as it is routinely done for diatomic molecules calculations. We showed that this term is well described by a sum of two-body terms. As mentioned before in ref.~\cite{pavolini1989}, it can be fitted by an exponential formula. Therefore, this term needs not to be calculated at each point of the three dimensional grid spanned by the degrees of freedom of a triatomic molecule on which the \emph{ab initio} calculations are performed. It can easily be added \emph{a posteriori} to cancel any non-physical behavior which might occur at short internuclear distances.

To estimate the quality of this basis set, we have compared our calculations to high level ones which uses a different method than ours, with the MOLPRO package. Such a comparison between different methods is crucial to assess their accuracy and their consistency, as no spectroscopic data are available for the Cs$_3$ system, as well as for most of the triatomic alkali systems. The calculations presented here are exploratory in nature: the small basis set used (Table~\ref{tab:basisSet}) makes the calculation time short enough to generate over a thousand \emph{ab initio} points. Figure~\ref{fig:cs2Triplet} shows the limitation of this basis set where the well depth of the triplet state of Cs$_{2}$ is overestimated by more than 100~cm$^{-1}$. For the quartet state of Cs$_{3}$, our calculated well depth is consequently 300~cm$^{-1}$ lower than the RCCSD(T) calculations. If we employ the basis set previously used for the caesium dimer~\cite{aymar2005} in the ECP+CPP-FCI method, this yields a very good agreement with the RCCSD(T) calculations, as the difference is now found at about $20\,\text{cm}^{-1}$ between them (Figure~\ref{fig:cs2Triplet}). The agreement is even more spectacular with the potential curve extracted from the spectroscopy of the lowest triplet state of Cs$_{2}$ \cite{xie2009}.

\begin{figure}[htbp]
		\begin{center}
			\includegraphics[width=15cm]{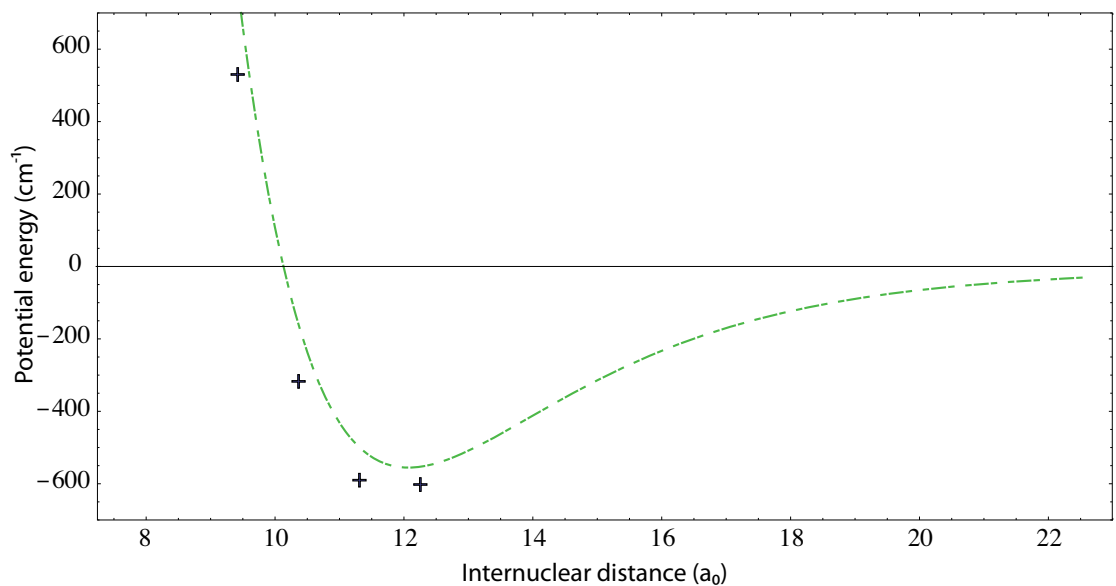}
		\end{center}
		\caption{Quartet $^{4}\Sigma_{u}^{+}$ state of Cs$_{3}$ in D$_{\infty h}$ symmetry. Green chain dotted curve : calculation at the RCCSD(T)/aug-ECP46MDF level of theory. Plus signs : preliminary calculations with the ECP+CPP-FCI approach using basis set “A” from Ref.~\cite{aymar2005}.}
		\label{fig:cs3Dinfh}
	\end{figure}

We are currently calculating a comprehensive set of \emph{ab initio} points with the ECP+CPP-FCI with this extended basis for the Cs$_{3}$ molecule which will be the subject of a future paper. Such FCI calculations now involve more than 500,000 Slater determinants. In Figure~\ref{fig:cs3Dinfh}, we display the first calculated points for the D$_{\infty h}$ symmetry of the lowest Cs$_{3}$ quartet state, compared to the MOLPRO calculation as described in Section~\ref{sec:preliminary_application_to_caesium_trimer}. The figure suggests that the results are in good agreement with each other as the position of the minimum looks similar. The point calculated around $12.3\,a_{0}$ with the ECP+CCP full CI method is deeper than the minimum of the MOLPRO curve by about $40\,\text{cm}^{-1}$, while the difference between corresponding triplet potential curves in Figure~\ref{fig:cs2Triplet} is seen to be about $20\,\text{cm}^{-1}$  This suggests that pairwise additivity of the forces is a reasonably good approximation in this case.
		
\section*{Acknowledgments} 
This work is performed in the framework of the network "Quantum Dipolar Molecular Gases" (QuDipMol) of the EUROCORES/EUROQUAM program of the European Science Foundation. J.D. acknowledges partial support of the French-German University (http://www.dfh-ufa.org), and R.G. support from {\it Institut Francilien de Recherches sur les atomes froids - IFRAF} (http://www.ifraf.org).

\newpage
\section*{References} 
\label{sec:references}

\begin{thebibliography}{10}

\bibitem{gerlich2008a}
D.~Gerlich,
\newblock {\em The study of cold collisions using ion guides and traps},
\newblock (in {\it Low temperatures and cold molecules}, p.121, ed. by I. W. M. Smith, World Scientific Publishing,  2008).

\bibitem{stwalley1999}
W.C. Stwalley and H.~Wang,
\newblock J. Molec. Spectrosc., {\bf 195},\hspace{0.25em}194  (1999).

\bibitem{jones2006}
K.~M. Jones, E.~Tiesinga, P.~D. Lett, and P.~S. Julienne,
\newblock Rev. Mod. Phys., {\bf 78},\hspace{0.25em}483  (2006).

\bibitem{meerakker2005}
S.~Y.~T. van~de Meerakker, N.~Vanhaecke, M.~P.~J. van~der Loo, G.~C.
  Groenenboom, and G.~Meijer,
\newblock Phys. Rev. Lett., {\bf 95},\hspace{0.25em}013003  (2005).

\bibitem{gilijamse2007}
J.~J. Gilijamse, S.~V. Hoekstra, S.~A. Meek, M.~Metsala, S.~Y.~T. van~de
  Meerakker, G.~Meijer, and G.~C. Groenenboom,
\newblock J. Chem. Phys., {\bf 127},\hspace{0.25em}221102  (2007).

\bibitem{demille2000}
D.~DeMille, F.~Bay, S.~Bickman, D.~Kawall, D.~Krause, Jr., S.~E. Maxwell, and
  L.~R. Hunter,
\newblock Phys. Rev. A, {\bf 61},\hspace{0.25em}052507  (2000).

\bibitem{hudson2002}
J.~J. Hudson, B.~E. Sauer, M.~R. Tarbutt, and E.~A. Hinds,
\newblock Phys. Rev. Lett., {\bf 89},\hspace{0.25em}023003  (2002).

\bibitem{kozlov2002}
M.~G. Kozlov and D.~DeMille,
\newblock Phys. Rev. Lett., {\bf 89},\hspace{0.25em}133001  (2002).

\bibitem{kawall2004}
D.~Kawall, F.~Bay, S.~Bickman, Y.~Jiang, and D.~DeMille,
\newblock Phys. Rev. Lett., {\bf 92},\hspace{0.25em}133007  (2004).

\bibitem{hinds1980}
E.~A. Hinds and P.~G.~H. Sandars,
\newblock Phys. Rev. A, {\bf 21},\hspace{0.25em}480  (1980).

\bibitem{cho1991}
D.~Cho, K.~Sangster, and E.~A. Hinds,
\newblock Phys. Rev. A, {\bf 44},\hspace{0.25em}2783  (1991).

\bibitem{ravaine2005}
B.~Ravaine, M.~G. Kozlov, and A.~Derevianko,
\newblock Phys. Rev. A, {\bf 72},\hspace{0.25em}012101  (2005).

\bibitem{veldhoven2004}
J.~van Veldhoven, J.~K\"upper, H.L. Bethlem, B.~Sartakov, A.J.A. van Roij, and
  G.~Meijer,
\newblock Eur. Phys. J. D, {\bf 31},\hspace{0.25em}337  (2004).

\bibitem{karr2005}
J.~Ph. Karr, S.~Kilic, and L.~Hilico,
\newblock J. Phys. B: At. Mol. Opt. Phys., {\bf
  38},\hspace{0.25em}853  (2005).

\bibitem{schiller2005}
S.~Schiller and V.~Korobov,
\newblock Phys. Rev. A, {\bf 71},\hspace{0.25em}032505  (2005).

\bibitem{chin2006}
C.~Chin and V.~V. Flambaum,
\newblock Phys. Rev. Lett., {\bf 96},\hspace{0.25em}230801  (2006).

\bibitem{demille2008}
D.~DeMille, S.~Sainis, J.~Sage, T.~Bergeman, S.~Kotochigova, and E.~Tiesinga,
\newblock Phys. Rev. Lett., {\bf 100},\hspace{0.25em}043202  (2008).

\bibitem{zelevinsky2008}
T.~Zelevinsky, S.~Kotochigova, and J.~Ye,
\newblock Phys. Rev. Lett., {\bf 100},\hspace{0.25em}043201  (2008).

\bibitem{hudson2006a}
E.~R. Hudson, H.~J. Lewandowski, B.~C. Sawyer, and J.~Ye,
\newblock Phys. Rev. Lett., {\bf 96},\hspace{0.25em}143004  (2006).

\bibitem{demille2002}
D.~DeMille,
\newblock Phys. Rev. Lett., {\bf 88},\hspace{0.25em}067901  (2002).

\bibitem{rabl2006}
P.~Rabl, D.~DeMille, J.~M. Doyle, M.~D. Lukin, R.~J. Schoelkopf, and P.~Zoller,
\newblock Phys. Rev. Lett., {\bf 97},\hspace{0.25em}033003  (2006).

\bibitem{kotochigova2006}
S.~Kotochigova and E.~Tiesinga,
\newblock Phys. Rev. A, {\bf 73},\hspace{0.25em}041405(R)  (2006).

\bibitem{rabl2007}
P.~Rabl and P.~Zoller,
\newblock Phys. Rev. A, {\bf 76},\hspace{0.25em}042308  (2007).

\bibitem{krems2005}
R.~V. Krems,
\newblock Int. Rev. Phys. Chem., {\bf 99},\hspace{0.25em}24  (2005).

\bibitem{barnett2006}
R.~Barnett, D.~Petrov, M.~Lukin, and E.~Demler,
\newblock Phys. Rev. Lett., {\bf 96},\hspace{0.25em}190401  (2006).

\bibitem{baranov2002}
M.~A. Baranov, M.~S. Mar\'enko, V.~S. Rychkov, and G.~V. Shlyapnikov,
\newblock Phys. Rev. A, {\bf 66},\hspace{0.25em}013606  (2002).

\bibitem{goral2002}
K.~Gor\'al, L.~Santos, and M.~Lewenstein,
\newblock Phys. Rev. Lett., {\bf 88},\hspace{0.25em}170406  (2002).

\bibitem{jaksch2002}
D.~Jaksch, V.~Venturi, J.~I. Cirac, C.~J. Williams, and P.~Zoller,
\newblock Phys. Rev. Lett., {\bf 89},\hspace{0.25em}040402  (2002).

\bibitem{donley2002}
E.~A. Donley, N.~R. Claussen, S.~T. Thompson, and C.~E. Wieman,
\newblock Nature, {\bf 417},\hspace{0.25em}529  (2002).

\bibitem{herbig2003}
J.~Herbig, T.~Kraemer, M.~Mark, T.~Weber, C.~Chin, H.-C. N\"{a}gerl, and
  R.~Grimm,
\newblock Science, {\bf 301},\hspace{0.25em}1510  (2003).

\bibitem{jochim2003}
S.~Jochim, M.~Bartenstein, A.~Altmeyer, G.~Hendl, C.~Chin, J.~Hecker Denschlag,
  and R.~Grimm,
\newblock Phys. Rev. Lett., {\bf 91},\hspace{0.25em}240402  (2003).

\bibitem{greiner2003}
M.~Greiner, C.~A. Regal, and D.~S. Jin,
\newblock Nature, {\bf 426},\hspace{0.25em}537  (2003).

\bibitem{zwierlein2003}
M.~W. Zwierlein, C.~A. Stan, C.~H. Schunck, S.~M.~F. Raupach, S.~Gupta,
  Z.~Hadzibabic, and W.~Ketterle,
\newblock Phys. Rev. Lett., {\bf 91},\hspace{0.25em}250401  (2003).

\bibitem{regal2003}
C.~A. Regal, C.~Ticknor, J.~L. Bohn, and D.~S. Jin,
\newblock Nature, {\bf 424},\hspace{0.25em}47  (2003).

\bibitem{bourdel2004}
T.~Bourdel, L.~Khaykovich, J.~Cubizolles, J.~Zhang, F.~Chevy, M.~Teichmann,
  L.~Tarruell, S.~J. J. M.~F. Kokkelmans, and C.~Salomon,
\newblock Phys. Rev. Lett., {\bf 93},\hspace{0.25em}050401  (2004).

\bibitem{bloch2005}
I.~Bloch,
\newblock J. Phys. B: At. Mol. and Opt. Phys., {\bf
  38},\hspace{0.25em}S629  (2005).

\bibitem{chin2004}
C.~Chin and R.~Grimm,
\newblock Phys. Rev. A, {\bf 69},\hspace{0.25em}033612  (2004).

\bibitem{bartenstein2004}
M.~Bartenstein, A.~Altmeyer, S.~Riedl, S.~Jochim, C.~Chin, J.~Hecker Denschlag,
  and R.~Grimm,
\newblock Phys. Rev. Lett., {\bf 92},\hspace{0.25em}120401  (2004).

\bibitem{zwierlein2004}
M.~W. Zwierlein, C.~A. Stan, C.~H. Schunck, S.~M.~F. Raupach, A.~J. Kerman, and
  W.~Ketterle,
\newblock Phys. Rev. Lett., {\bf 92},\hspace{0.25em}120403  (2004).

\bibitem{bethlem2003}
H.~L. Bethlem and G.~Meijer,
\newblock Int. Rev. Phys. Chem., {\bf 22},\hspace{0.25em}73  (2003).

\bibitem{hutson2006}
J.~M. Hutson and P.~Sold\'an,
\newblock Int. Rev. Phys. Chem., {\bf 25},\hspace{0.25em}497  (2006).

\bibitem{kohler2006}
T.~K\"ohler, K.~G\'{o}ral, and P.~S. Julienne,
\newblock Rev. Mod. Phys., {\bf 78},\hspace{0.25em}1311  (2006).

\bibitem{viteau2008}
M.~Viteau, A.~Chotia, M.~Allegrini, N.~Bouloufa, O.~Dulieu, D.~Comparat, and
  P.~Pillet,
\newblock Science, {\bf 321},\hspace{0.25em}232  (2008).

\bibitem{deiglmayr2008a}
J.~Deiglmayr, A.~Grochola, M.~Repp, K.~M\"{o}rtlbauer, C.~Gl\"{u}ck, J.~Lange,
  O.~Dulieu, R.~Wester, and M.~Weidem\"{u}ller,
\newblock Phys. Rev. Lett., {\bf 101},\hspace{0.25em}133004  (2008).

\bibitem{danzl2008}
J.~G. Danzl, E.~Haller, M.~Gustavsson, M.~J. Mark, R.~Hart, N.~Bouloufa,
  O.~Dulieu, H.~Ritsch, and H.-C. N\"{a}gerl,
\newblock Science, {\bf 321},\hspace{0.25em}1062  (2008).

\bibitem{mark2009}
M.~J. Mark, J.~G. Danzl, E.~Haller, M.~Gustavsson, N.~Bouloufa, O.~Dulieu,
  H.~Salami, T.~Bergeman, H.~Ritsch, R.~Hart, and H.-C. N\"agerl,
\newblock arXiv:0811.0695.

\bibitem{ni2008}
K.-K. Ni, S.~Ospelkaus, M.~H.~G. de~Miranda, A.~Peer, B.~Neyenhuis, J.~J.
  Zirbel, S.~Kotochigova, P.~S. Julienne, D.~S. Jin, and J.~Ye,
\newblock Science, {\bf 322},\hspace{0.25em}231  (2008).

\bibitem{lang2008}
F.~Lang, P.~v.~d. Straten, B.~Brandst\"{a}tter, G.~Thalhammer, K.~Winkler,
  P.~S. Julienne, R.~Grimm, and J.~Hecker Denschlag,
\newblock Nature Physics, {\bf 4},\hspace{0.25em}223  (2008).

\bibitem{staanum2006}
P.~Staanum, S.~D. Kraft, J.~Lange, R.~Wester, and M.~Weidem\"{u}ller,
\newblock Phys. Rev. Lett., {\bf 96},\hspace{0.25em}023201  (2006).

\bibitem{zahzam2006}
N.~Zahzam, T.~Vogt, M.~Mudrich, D.~Comparat, and P.~Pillet,
\newblock Phys. Rev. Lett., {\bf 96},\hspace{0.25em}023202  (2006).

\bibitem{chin2005}
C.~Chin, T.~Kraemer, M.~Mark, J.~Herbig, P.~Waldburger, H.-C. N\"agerl, and
  R.~Grimm,
\newblock Phys. Rev. Lett., {\bf 94},\hspace{0.25em}123201  (2005).

\bibitem{hudson2008}
E.~R. Hudson, N.~B. Gilfoy, S.~Kotochigova, J.~M. Sage, and D.~DeMille,
\newblock Phys. Rev. Lett., {\bf 100},\hspace{0.25em}203201  (2008).

\bibitem{soldan2003}
P.~Sold\'an, M.~T. Cvita\v{s}, and J.~M. Hutson,
\newblock Phys. Rev. A, {\bf 67},\hspace{0.25em}054702  (2003).

\bibitem{colavecchia2003a}
F.~D. Colavecchia, J.~P.~Burke Jr., W.~J. Stevens, M.~R. Salazar, G.~A. Parker,
  and R.~T Pack,
\newblock J. Chem. Phys., {\bf 118},\hspace{0.25em}5484  (2003).

\bibitem{cvitas2007}
M.~T. Cvita\v{s}, P.~Sold\'an, J.~M. Hutson, P.~Honvault, and J.-M. Launay,
\newblock J. Chem. Phys., {\bf 127},\hspace{0.25em}074302  (2007).

\bibitem{higgins2000}
J.~Higgins, T.~Hollebeek, J.~Reho, T.-S. Ho, K.~K. Lehmann, H.~Rabitz,
  G.~Scoles, and M.~Gutowski,
\newblock J. Chem. Phys., {\bf 112},\hspace{0.25em}5751  (2000).

\bibitem{simoni2009}
A.~Simoni, J.-M. Launay, and P.~Sold\'an,
\newblock arXiv:0901.3129  (2009).

\bibitem{quemener2005}
G.~Qu\'em\'ener, P.~Honvault, J.-M. Launay, P.~Sold\'an, D.~E. Potter, and
  J.~M. Hutson,
\newblock Phys. Rev. A, {\bf 71},\hspace{0.25em}032722  (2005).

\bibitem{soldan2008u}
P.~Sold\'an,
\newblock unpublished results.

\bibitem{soldan2002}
P.~Sold\'{a}n, M.~T. Cvita\v{s}, J.~M. Hutson, P.~Honvault, and J.-M. Launay,
\newblock Phys. Rev. Lett., {\bf 89},\hspace{0.25em}153201  (2002).

\bibitem{quemener2004}
G.~Qu\'em\'ener, P.~Honvault, and J.~M. Launay,
\newblock Eur. Phys. J. D, {\bf 30},\hspace{0.25em}201  (2004).

\bibitem{cvitas2005a}
M.T. Cvita\v{s}, P.~Sold\'an, J.~M. Hutson, P.~Honvault, and J.-M. Launay,
\newblock Phys. Rev. Lett., {\bf 94},\hspace{0.25em}033201  (2005).

\bibitem{cvitas2005}
M.~T. Cvita\v{s}, P.~Sold\'an, J.~M. Hutson, P.~Honvault, and J.-M. Launay,
\newblock Phys. Rev. Lett., {\bf 94},\hspace{0.25em}200402  (2005).

\bibitem{quemener2007}
G.~Qu\'{e}m\'{e}ner, J.-M. Launay, and P.~Honvault,
\newblock Phys. Rev. A, {\bf 75},\hspace{0.25em}050701  (2007).

\bibitem{li2008a}
X.~Li, G.~A. Parker, P.~Brumer, I.~Thanopulos, and M.~Shapiro,
\newblock Phys. Rev. Lett., {\bf 101},\hspace{0.25em}043003  (2008).

\bibitem{li2008b}
X.~Li, G.~A. Parker, P.~Brumer, I.~Thanopulos, and M.~Shapiro,
\newblock J. Chem. Phys., {\bf 128},\hspace{0.25em}124314  (2008).

\bibitem{li2008c}
X.~Li and G.~A. Parker,
\newblock J. Chem. Phys., {\bf 128},\hspace{0.25em}184113  (2008).

\bibitem{hutson2007}
J.~M. Hutson and P~Sold\'an,
\newblock Int. Rev. Phys. Chem., {\bf 26},\hspace{0.25em}1  (2007).

\bibitem{soldan2008}
P.~Sold\'an,
\newblock Phys. Rev. A, {\bf 77},\hspace{0.25em}054501  (2008).

\bibitem{aymar2005}
M.~Aymar and O.~Dulieu,
\newblock J. Chem. Phys., {\bf 122},\hspace{0.25em}204302  (2005).

\bibitem{aymar2006}
M.~Aymar, O.~Dulieu, and F.~Spiegelmann,
\newblock J. Phys. B: At. Mol. and Opt. Phys., {\bf
  39},\hspace{0.25em}S905  (2006).

\bibitem{deiglmayr2008}
J.~Deiglmayr, M.~Aymar, R.~Wester, M.~Weidem\"uller, and O.~Dulieu,
\newblock J. Chem. Phys., {\bf 129},\hspace{0.25em}064309  (2008).

\bibitem{aymar2009}
M.~Aymar, J.~Deiglmayr, and O.~Dulieu,
\newblock Can. J. Phys, {\bf in print},  (2009).

\bibitem{huron1973}
B.~Huron, J.-P. Malrieu, and P.Rancurel,
\newblock J. Chem. Phys., {\bf 58},\hspace{0.25em}5745  (1973).

\bibitem{durand1974}
P.~Durand and J.C. Barthelat,
\newblock Chem. Phys. Lett., {\bf 27},\hspace{0.25em}191  (1974).

\bibitem{durand1975}
P.~Durand and J.C. Barthelat,
\newblock Theor. chim. Acta, {\bf 38},\hspace{0.25em}283  (1975).

\bibitem{muller1984}
W.~M\"uller and W.~Meyer,
\newblock J. Chem. Phys., {\bf 80},\hspace{0.25em}3311  (1984).

\bibitem{foucrault1992}
M.~Foucrault, Ph. Milli\'e, and J.P. Daudey,
\newblock J. Chem. Phys., {\bf 96}(2),\hspace{0.25em}1257  (1992).

\bibitem{spiegelmann1989}
F.~Spiegelmann, D.~Pavolini, and J.-P. Daudey,
\newblock J. Phys. B, {\bf 22},\hspace{0.25em}2465  (1989).

\bibitem{pavolini1989}
D.~Pavolini, T.~Gustavsson, F.~Spiegelmann, and J.-P. Daudey,
\newblock J. Phys. B, {\bf 22},\hspace{0.25em}1721  (1989).

\bibitem{magnier1993}
S.~Magnier, P.~Milli\'e, O.~Dulieu, and F.~Masnou-Seeuws,
\newblock J. Chem. Phys., {\bf 98},\hspace{0.25em}7113  (1993).

\bibitem{magnier1996}
S.~Magnier and P.~Milli\'e,
\newblock Phys. Rev. A, {\bf 54},\hspace{0.25em}204  (1996).

\bibitem{MOLPRO_brief}
H.-J. Werner, P.~J. Knowles, R.~Lindh, F.~R. Manby, M.~{Sch\"{u}tz}, et~al.,
\newblock Molpro, version 2006.1, a package of ab initio programs, (2006),
\newblock see http://www.molpro.net.

\bibitem{iron2003}
M.~Oren M.~A.~Iron and J.~M.~L. Martin,
\newblock Mol. Phys., {\bf 101},\hspace{0.25em}1345  (2003).

\bibitem{lim2005a}
B.~Metz I.~S.~Lim, P.~Schwerdtfeger and H.~Stoll,
\newblock J. Chem. Phys., {\bf 122},\hspace{0.25em}104103  (2005).

\bibitem{knowles1993}
C.~Hampel P.~J.~Knowles and H.~J. Werner,
\newblock J. Chem. Phys, {\bf 99},\hspace{0.25em}5219  (1993).

\bibitem{cizek1966}
J.~\v{C}\'\i\v{z}ek,
\newblock J. Chem. Phys., {\bf 45},\hspace{0.25em}4526  (1966).

\bibitem{boys1970}
S.~F. Boys and F.~Bernardi,
\newblock Mol. Phys., {\bf 19},\hspace{0.25em}553  (1970).

\bibitem{jeung1997}
G-H. Jeung,
\newblock J. Molec. Spect., {\bf 182},\hspace{0.25em}113  (1997).

\bibitem{coker1976}
H.~Coker,
\newblock J. Phys. Chem., {\bf 80},\hspace{0.25em}2073  (1976).

\bibitem{xie2009}
X.~Xie {\it et al},
\newblock submitted  (2009).

\end{thebibliography}

\end{document}